\documentclass[conference]{IEEEtran}
\IEEEoverridecommandlockouts
\usepackage{cite}
\usepackage{amsmath,amssymb,amsfonts}
\usepackage{algorithmic}
\usepackage{graphicx}
\usepackage{textcomp}
\usepackage{xcolor}
\usepackage{subcaption}
\usepackage{enumerate}
\def\BibTeX{{\rm B\kern-.05em{\sc i\kern-.025em b}\kern-.08em
    T\kern-.1667em\lower.7ex\hbox{E}\kern-.125emX}}

\newcommand{\oprocendsymbol}{\hbox{$\bullet$}}
\newcommand{\oprocend}{\relax\ifmmode\else\unskip\hfill\fi\oprocendsymbol}

\newcommand{\diag}{\operatorname{diag}}
\newcommand{\tp}{^{\text{\scalebox{0.8}{$\top$}}}}

\newcommand{\defin}{:=}

\newcommand{\until}[1]{\{1,\dots,#1\}}

\newcommand{\realnonnegative}{{\mathbb{R}}_{\ge 0}}
\newcommand{\real}{{\mathbb{R}}}

\newcommand{\complex}{{\mathbb{C}}}

\newcommand{\norm}[1]{\|#1\|_{\text{\scalebox{1}{$2$}}}}

\newcommand{\absolute}[1]{|#1|}

\newcommand{\ones}{\mathrm{1}}

\newcommand{\asensor}[1]{A^{(#1)}}

\newcommand{\stsensor}[1]{a^{(#1)}}
\newcommand{\ysensor}[1]{y^{(#1)}}

\newcommand{\xsensor}[1]{x^{(#1)}}
\newcommand{\xsensormean}{\bar{x}}
\newcommand{\ssensor}[1]{s^{(#1)}}

\newcommand{\msen}[1]{m_{#1}}

\newcommand{\noisesensor}[1]{w^{(#1)}}

\newcommand{\grouplabelset}{\mathcal{G}}
\newcommand{\ngroups}{\text{\scalebox{0.8}{\# groups}}}
\newcommand{\grouplast}{g_\ngroups}

\newcommand{\normgroup}[1]{\|#1\|_{\text{\scalebox{1}{$\grouplabelset$,2,1}}}}
\newcommand{\normlasso}[1]{\|#1\|_{\text{\scalebox{1}{2,1}}}}

\newcommand{\add}{a_\text{\scalebox{0.8}{DD}} }
\newcommand{\adr}{a_\text{\scalebox{0.8}{DR}} }
\newcommand{\ard}{a_\text{\scalebox{0.8}{RD}} }
\newcommand{\arr}{a_\text{\scalebox{0.8}{RR}} }

\newcommand{\nh}{{N}}
\newcommand{\naz}{{n_\phi}}
\newcommand{\nrho}{{n_\rho}}
\newcommand{\nhprime}{{n_h}}
\newcommand{\ysub}[1]{y_{\text{\scalebox{0.8}{($#1$)}}}}
\newcommand{\xsub}[1]{x_{\text{\scalebox{0.8}{($#1$)}}}}
\newcommand{\Asub}[1]{A_{\text{\scalebox{0.8}{($#1$)}}}}
\newcommand{\ysubhor}[1]{y_{\text{\scalebox{0.8}{($#1$)}}}^{\text{\scalebox{0.8}{Hor}}}}
\newcommand{\Asubhor}[1]{A_{\text{\scalebox{0.8}{($#1$)}}}^{\text{\scalebox{0.8}{Hor}}}}

\newcommand{\phimin}{\phi_\text{\scalebox{0.9}{min}}}
\newcommand{\phimax}{\phi_\text{\scalebox{0.9}{max}}}

\newcommand{\thetamax}{\theta_\text{\scalebox{0.9}{max}}}
\newcommand{\hmin}{h_\text{\scalebox{0.9}{min}}}
\newcommand{\hmax}{h_\text{\scalebox{0.9}{max}}}

\newcommand{\xazimap}{\bar{x}^\text{\scalebox{0.8}{Az}}}
\newcommand{\xheightmap}{\bar{x}^\text{\scalebox{0.8}{Height}}}

\begin{document}

\title{Height estimation for automotive MIMO radar\\ with group-sparse reconstruction
\thanks{All authors contributed equally, and are with the department of Cognitive Radar at Fraunhofer FHR, Wachtberg, Germany.}
}

\author{
	\IEEEauthorblockN{Renato Simoni}
	%
	\and
		\IEEEauthorblockN{ David Mateos-N\'u\~nez}
	%
	\and
	\IEEEauthorblockN{Mar\'ia A. Gonz\'alez-Huici}
	%
	\and
	\IEEEauthorblockN{Aitor Correas-Serrano}
	%
}

\maketitle

\begin{abstract}
A method is developed for sequential azimuth and height estimation of small objects at far distances in front of a moving vehicle using coherent or mutually incoherent MIMO arrays. The model considers phases and amplitudes of superposition of near-field multipath signals produced by specular non-diffusive ground-reflections. The reflection phase shift and power attenuation due to the interaction with the ground is assumed unknown and is estimated jointly. Group-sparsity allows combining measurements along the trajectory of the vehicle provided that the road is flat as well as measurements from multiple incoherent sensors at different locations in the vehicle. 
 This model can be formulated for non-uniform sparse arrays in 2D and 3D with subsets of antennas at the same height and the resulting inverse problem can be approximated with efficient methods such as Block Orthogonal Matching Pursuit.
 It is shown in simulations that the proposed approach significantly increases estimation accuracy and decreases false alarms, both crucial for the detection of small objects at far distances.
%
%
\end{abstract}


\section{Introduction}

Radar systems for height and azimuth estimation have crucial importance for collision avoidance systems in motor-vehicles, because they work in most weather and lighting conditions, allow radial velocity estimation via Doppler measurements, and can provide high spatial resolution at ranges up to $200$ m. In particular, at high speeds, collision avoidance decisions need to be made at distances larger than $80$ m regarding the presence of obstacles. This requires an estimation of the dimensions of small objects on the road at long distances, e.g. at distances between $80$ m and $150$ m, where radar data can be collected and processed to determine if the obstacle can be driven over or needs to be avoided.

The problem of height estimation in the context of automotive radar up to now has been addressed in two scenarios. A first scenario is the detection of small objects, e.g. curbs, or classification of medium objects, e.g., poles, at short distances of $1$ to $5$ m, related to the problem of parking~\cite{SO-CW:17,AL-MH-JD-CW:17}. The second scenario is the detection of high objects like bridges at far distances of $50$ to $200$ m, related to the problem of classification of objects as in-path versus over-pass~\cite{FD-JK-FS-JD-KD:11,FE-MW-PH:17}, and the height estimation of gates,~\cite{AL-MH-JD-CW:17}. 

%
 The method in~\cite{FD-JK-FS-JD-KD:11} uses the fact that the road-reflected multi-path signals interfere with one another, and the power received is periodic in the change of variable of inverse distance to obstacle~\cite[pp. 451]{SMK:81}. This period is related to the height of a single scatterer and can be estimated extracting the peaks of the power spectral density using Fast Fourier Transforms (FFTs) or other methods.
 %
  The reasons this method is insufficient for estimating the height of low objects are the following: i) the number of cycles of the interference pattern is even less than one for small and distant targets, making the power spectral density estimation a challenge even for parametric models like Burg method; ii) the DC component (or average component) introduces a disturbance because it is unknown; and iii) a second scatterer changes the relationship of frequency to height, in other words, the model is highly dependent on the number of scatterers. The fact that this method uses amplitude or power information and not phases across an array of antennas is an important caveat for extensions using sparse reconstruction because there is no linear super-position of signals for multiple scatterers.
The method in~\cite{FE-MW-PH:17} uses maximum likelihood estimates and hypothesis-testing for model selection between three categories, one or two targets, and a one-target model in the presence of multi-path. The model used is far-field and although it can be generalized, the concentrated log-likelihood~\cite{PE-KCS:90} depends on parameters like the attenuation coefficient that need to be estimated using costly multi-dimensional optimization.
Moreover, this work does not combine measurements along the trajectory of the vehicle or from different sensors, and a single measurement may be insufficient using an array of practical number of elements and dimensions for automotive radar. 
Other principles for height estimation used for height estimation with automotive radar are summarized in
Table~\ref{table:literature-height-estimation}.

\begin{table}[t]
	\centering
	\caption{Summary of observation models used in the literature of automotive radar height estimation}
	\label{table:literature-height-estimation}
	\begin{tabular}{ p{1cm}| p{3.0cm} | p{3cm} }
		\hline
		\noalign{\vskip 1pt} 
		Work & Principle & Application
		\\[1pt]
		\hline
		\hline
		\noalign{\vskip 1pt} 
		\cite{FD-JK-FS-JD-KD:11}&
		 Envelope of interference signal across distances & 
		 \textbf{Bridges:} Classification as over-pass versus in-path
		\\[2pt]
		\cite{SO-CW:17}&
		 Envelope of interference signal across array & 
		\textbf{Parking:} Classification of small objects (curbs, cans, poles) at short distances
		\\[2pt]
		\cite{AL-MH-JD-CW:17,AL-MH-JD-CW:17b}&
		Delay difference between paths & 
		\textbf{Parking:} Estimation of height at close distances
		 \\[2pt]
		\cite{FE-MW-PH:17}&
	DoA estimation of target and mirror image and averaging over distance & 
	\textbf{High targets:} Estimation of height of objects of above $1$-$3$ m at far distances
		\\
		\hline
		\hline
		\noalign{\vskip 2pt}
	\end{tabular}
\end{table}


Group-sparse reconstruction has been used for radar signal processing, including
incoherent sensor fusion in multi-static ISAR~\cite{SB-JE:15}, passive radar networks~\cite{MW:16}, 
tracking~\cite{LG-YDZ-QW-MGA-BH:16}, 
%
%
%
and
time-frequency estimation~\cite{YDZ-LG-QW-MGA:15}. 
 In other work of the authors~\cite{DMN-MGH-RS-ACS:19}, a design principle based on group-sparse reconstruction is exploited for optimization of antenna positions of mutually incoherent apertures with the effect of sidelobe averaging while maintaining a thin mainlobe with a most efficient use of space available.

To our knowledge this is the first contribution using a sequential group-sparse reconstruction method for azimuth and height estimation using near-field multi-path models combining measurements from multiple points along the trajectory and possibly multiple sensors to enhance the estimation, and capable of dealing with an unknown reflection attenuation. This work is part of a patent application~\cite{patent_height_est:18}.


%
%

\textit{Notations:} We denote by $\real^{m\times n}$ and $\complex^{m\times n}$ the set of real and complex matrices, and use $(.)\tp$ and $(.)^H$ to denote the transpose and conjugate transpose, respectively. $\ones_n\in\real^{n\times 1}$ represents the column vector of ones. The entry-wise absolute value or complex modulus of a vector $x$ is denoted by $\absolute{x}$, and $x[1:n: \text{end}]$ forms the vector with every $n$-th entry of $x$.

\textit{Organization:} 
In Section~\ref{sec-prelimns-doa-group-sparse} we review the group-sparse reconstruction approach for angular estimation using multiple measurements. In Section~\ref{sec-height-estimation-model}, we extend this formulation for multi-path models with unknown reflection coefficient, and define an approach where  measurements are combined incoherently for an  interval along the trajectory and/or using multiple sensors. We present analyses with synthetic measurements in Section~\ref{sec-simulation results}, including design recommendations about the position of the sensor and their geometry. Section~\ref{sec-conclusions} concludes with remarks and ideas for future work.

\section{Preliminaries on group-sparse reconstruction for DoA estimation}\label{sec-prelimns-doa-group-sparse}

Here we introduce the DoA estimation model for a set of mutually incoherent array signals. This is employed in our formulation when we estimate the height using incoherent fusion of measurements at multiple points of the trajectory of the vehicle, and optionally using multiple incoherent sensors. 
%


\subsection{Signal model for mutually incoherent array signals}\label{subsec:signal-model}
\begin{figure}[]
	\hspace*{0.10cm}
	\includegraphics[width=0.97\linewidth]{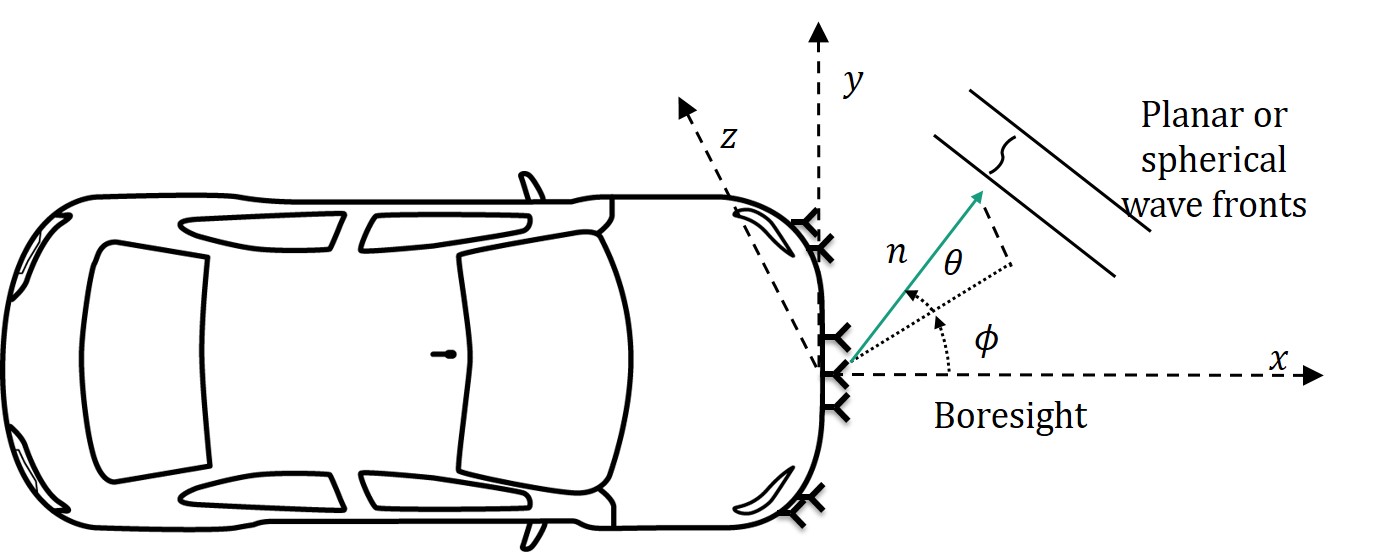}

	\caption{Sets of antennas in a reference frame.
	A random initial phase is assumed between mutually incoherent array apertures.
		%
	}
	\label{fig:diagram-incoherent-apertures}
\end{figure}

Next we describe the model 
for general parameter estimation using
mutually incoherent array signals, i.e., each with a random initial phase. We use this model in two scenarios: i)~modeling of radar signals for a single array sensor during a time-span that exceeds the coherent processing interval and/or at different points along the trajectory of the vehicular radar, and ii)~the case of mutually incoherent radar array sensors to account for lack of synchronization or deformations between widely separated antenna elements, cf.~Fig.~\ref{fig:diagram-incoherent-apertures}.

We assume a 1-snapshot model after match-filtering or preliminary 
range-Doppler processing. 
%
%
A coherent array measurement $l\in\until{L}$ for a sum of $K$ sources, each with parameter $\phi_k$, is modeled by
\begin{align}\label{eq:signal-model-multisensors}
\ysensor{l}=\sum_{k=1}^{K}\ssensor{l}_k \stsensor{l}(\phi_k)+\noisesensor{l}  .
\end{align}
where $\ssensor{l}_k\in\complex$ is the complex amplitude of source~$k$, accounting for path propagation loss, target RCS, processing gain of matched-filtering, and a random initial phase associated to coherent aperture~$l$;  and $\noisesensor{l}\in\complex^{\msen{l}}$ is the noise after matched-filtering.
%

Of particular interest in this work is the near-field model for angular estimation. The reason of choosing this model over the far-field case is that the phases of the super-position of the signals produced by reflections on the ground are modeled more accurately considering the geometric distances of each path.
The range $r$ is assumed previously estimated and indeed only an approximation is required to evaluate a model of the phase differences between antennas. The intuition is that in the extreme case of the far-field model, the exact distance to the target only introduces an initial phase that can be factored out for all the antennas and is not required for angular estimation. However, the initial phase resulting from each path plays a role when modeling the super-position of ground-reflected paths that have similar length for low objects. That is, we do not require an exact range estimate, but for a given range we require consistency in the relative distances between paths.

 Following~\cite{AME-TET:15}, 
 for a set of antennas  $\{p_i=[x_i,y_i,z_i]\tp\}$, and a point narrow-band source at azimuth and elevation $\xi=[\phi,\theta]$, and range~$r$, the geometric near-field phases are modeled by
\begin{align}\label{eq:near-field-steering}
[a_{Rx}(\xi,r)]_i\defin\exp\bigg(j \tfrac{2\pi}{\lambda} r
\sqrt{1-\tfrac{2}{r} n\tp p_i
	+ \tfrac{p_i\tp p_i}{r^2}}\,\, \bigg) ,
\end{align}
where $n\defin[\cos(\theta)\cos(\phi), \cos(\theta)\sin(\phi), \sin(\theta)]$.
By reciprocity of transmission and reception, the MIMO steering vector upon matched filtering is 
\begin{align} \label{eq:virtual-steering}
a_{\text{Virt}}(\xi,r)\defin a_{Tx}(\xi,r)\otimes a_{Rx}(\xi,r).
\end{align}
%
Model~\eqref{eq:signal-model-multisensors} does not assume that relative phases are measured between different points of the trajectory of the vehicle or among mutually incoherent sensors. 
In the optional case of multiple incoherent sensors, the relative positions of antennas \textit{within} each coherent aperture must be known accurately in relation to the wavelength, 
but distances between apertures can be approximated, and deformations and vibrations proportional to the distance between the apertures do not have impact because inter-aperture phases are not used.

\subsection{Review of group-sparse reconstruction}\label{subsec:gs-reconstruction}

Next we formulate the parameter estimation problem for the signal model~\eqref{eq:signal-model-multisensors} as a group-sparse reconstruction problem, which entails fitting each coherent measurement (for one sensor at multiple times or for multiple sensors) as a sparse linear super-position of sources in a grid of hypotheses, leveraging the correspondence between observations at the multiple time instants or sensors. Formally, one way of approximating the reconstruction coefficients is to solve the optimization
\begin{align}\label{eq:mininimization-group-sparse-reconstruction}
\min_{\xsensor{1},...,\xsensor{L}}\sum_{l=1}^{L}
\alpha_l\norm{\ysensor{l}-\asensor{l}\xsensor{l}}^2 +\alpha_0\normlasso{X},
\end{align}
where the sensing matrix
\begin{align}\label{eq:sensing-matrix-azimuth}
\asensor{l}\defin [\stsensor{l}(\phi_1) \cdots \stsensor{l}(\phi_N)]\in\complex^{\msen{l}\times N}
\end{align}
contains as columns the measurement model for coherent signal~$l$, e.g.,~\eqref{eq:virtual-steering}, evaluated in a grid $\{\phi_1,\dots, \phi_N\}$ for the relevant FoV. To introduce the model, we have considered azimuth estimation assuming that elevation is $0$. The sequential estimation of azimuth and elevation is presented in Section~\ref{sec-height-estimation-model}.

The matrix $X\defin[\xsensor{1} \cdots \xsensor{L}]\in\complex^{N\times L} $ contains as columns the coefficients for the reconstruction of each of the coherent signals, and the support is induced to be common across columns with sparsity induced among rows (hypotheses), via the 2-1 block-norm,


\begin{align}\label{eq:group-sparsity-constraint}
	\normlasso{X}\defin\sum_{i=1}^N \norm{x[i]} ,
\end{align}
where $x[i]\defin[\xsensor{1}_i \dots \xsensor{L}_i]\in\complex^L$, the $i$th row of $X$, collects the reconstruction coefficients for hypothesis $i$, i.e., and its absolute value is the amplitude of hypotheses $i$ in each of the coherent measurements $l\in\{1,...,L\}$.
%
%
The coefficient $\alpha_0>0$ is chosen big when fewer targets are expected, and 
$\alpha_1,...,\alpha_L>0$  ponderate the confidence level (variance) of each coherent measurement.

Problem~\eqref{eq:mininimization-group-sparse-reconstruction} is related to Basis Pursuit denoising (BPDN)~\cite{SSC-DLD-MAS:98, EVDB-MPF:08, spgl1:2007}, but one can alternatively use greedy methods like BOMP~\cite{TB-MED:08,YCE-PK-HB:10} (cf.~\cite{SB-JE:15} that in particular addresses model~\eqref{eq:mininimization-group-sparse-reconstruction}), and others iterative methods for sparse signals in transformed domains as Iterative Method with Adaptive Thresholding (IMAT)~\cite{FM-MA-PI-PP-SJ-AG:12, AA-MA-JT-FM:16}.
 
The method BOMP instead of specification of $\{\alpha_i\}$, requires a stopping criterion based on number of targets or final size of the estimation residual compared to the assumed noise levels.

To estimate the DoAs from the reconstruction coefficients, one option is to define the average of the moduli of the reconstruction coefficients for each hypothesis index $i\in\until{N}$ across all coherent measurements,  
\begin{align}\label{eq:doa-reconstruction-map}
	\xsensormean(i) \defin \sum_{l=1}^{L} \absolute{\xsensor{l}_i}.
\end{align}
We refer to the quantity $\xsensormean\in\realnonnegative^N$ as the DoA map (in analogy to the range-Doppler map although the differences are that here sparsity has been induced, and the domain can be spatial frequency, e.g., in the case of partial Fourier matrix, or an arbitrary transform domain, hence the modeling flexibility). Each entry $i\in\until{N}$ of $\xsensormean$ with a magnitude exceeding a threshold $ \xsensormean(i) > \gamma$  constitutes a \textit{declared} target, or declaration, with DoA estimate $\phi_i$.  
 The threshold~$\gamma$ 
 can depend on the array, and SNR,
 and can be found with training data of real or simulated scenarios, and optimized for each array.
%
%


Note that~\eqref{eq:group-sparsity-constraint} requires a correspondence between hypotheses across all mutually incoherent measurement models, which requires a common reference frame. This is specially relevant in the scenarios described at the beginning of the section, namely, for data integration over a long trajectory of the vehicle, where inertial systems may be necessary to adjust hypotheses of the target parameters, or, in the case of array measurements with separated mutually incoherent sensors with different orientations, cf.~Fig.~\ref{fig:diagram-incoherent-apertures}, where, in particular, some hypotheses are not present in both field of views. 

\subsection{Group-sparsity for arbitrary partition}\label{subsec:gs-general-partition}
Group sparsity models include more general settings than described above of incoherent processing of multiple array measurements. Consider a linear system $y\approx Ax$, with $A\in\complex^{M\times N}$. Define a partition of the column indexes in the sensing matrix~$A$, and associated entries of $x$, among sets called groups $\grouplabelset=\{g_1,...,\grouplast \}$, and apply a penalty or constraint in terms of the group- or block-norm defined by
 \begin{align}\label{eq:def-group-norm-general}
\normgroup{x}\defin\sum_{l=1}^{\ngroups} \norm{x_{\{k\in g_l\}}} ,
 \end{align}
where $\{k\in g_l\}$ represents the indexes in group $g_l$. A group-sparse vector $x$ with sparsity level $K$ is one where entries are nonzero only if they have indexes in a set of groups $g_{i_1}$,...,$g_{i_K}$. 
Again, such sparse reconstructions can be found using BPDN~\cite{spgl1:2007},  BOMP~\cite{YCE-PK-HB:10} and in some special cases IMAT~\cite{AA-MA-JT-FM:16}.
A convenient way of expressing a partition is to define a vector of labels
\begin{align}\label{eq:label_vector}
v_\grouplabelset\defin[v_1,...,v_N]
\end{align}
with $v_k=l \Leftrightarrow  k\in g_l$.

The intuition is that if hypotheses are competitive/exclusive, we wish to penalize their amplitude with the 1-norm to induce sparsity, i.e., they should be in different groups and therefore have associated different labels, whereas if hypotheses  can be reconciled or are complementary, we penalize their amplitude with the 2-norm, which does not induce sparsity. We use this notion, e.g., to express the possibility that the reflection coefficient can vary between measurements, while hypotheses of height are exclusive because height is constant.

\section{Height estimation model with group-sparse reconstruction}\label{sec-height-estimation-model}

In previous section we have reviewed the group-sparse reconstruction problem for a single parameter, e.g., azimuth, in a scenario where there is a set of mutually incoherent measurements, i.e., the phases cannot be measured between subsets of coherent (but mutually incoherent) antennas. The model of mutual incoherence is important to combine measurements at multiple points of the trajectory of the vehicle and from multiple sensors.
 In the next section we describe the model of near-field multi-path superposition, and the sequential procedure for azimuth and height estimation.

%

\begin{figure}[]
		\includegraphics[width=0.99\linewidth]{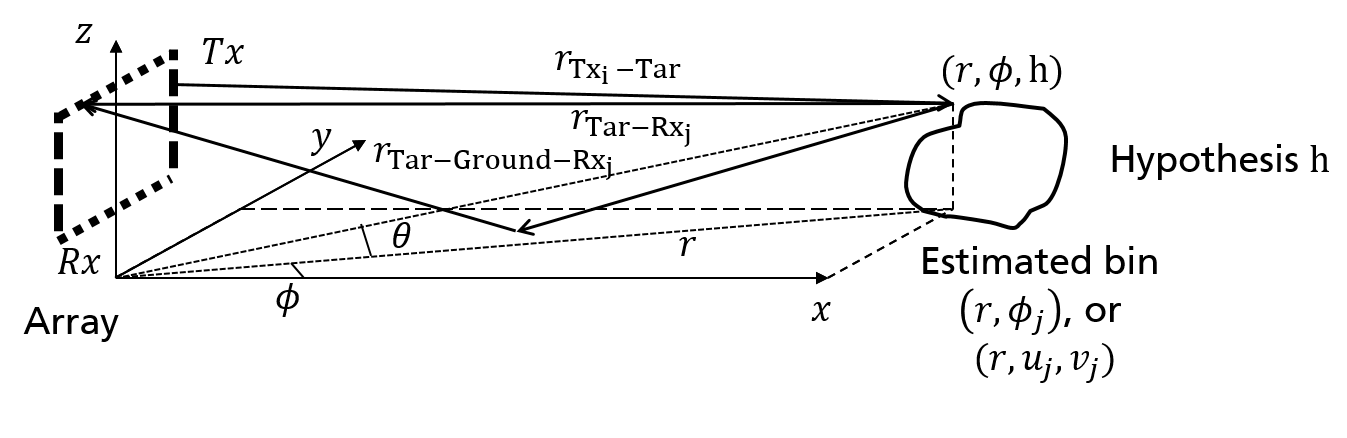}
		\caption{Obstacle in a road and the direct and ground-reflected paths from a given Tx to a scatterer, and back to a given Rx.
			For estimation of height, we consider hypotheses of height at the previously estimated azimuth bins. In cases~(i) and~(iii) the starting estimates are the range and azimuth bins $(r,\phi_j)$, and in case~(ii) the range and the spatial frequency bins in the depth and horizontal axes $(r,u_j, u_j)$.
		}
		\label{fig:model-4path}

\end{figure}

\subsection{Near-field multi-path model}\label{subsec:multipath-steering-vector}

The estimation of height is more accurate if we leverage the multi-path contributions provided that the road is flat and the reflection is not diffusive, i.e., there is a single reflection point on the road, cf.~Fig.~\ref{fig:model-4path}.
This is because the angle of the direct path from the target and the angle of the reflected path are related to the height.

The multi-path steering vector is defined as the super-position of the $4$ paths for a hypothetical reflection coefficient $\rho\in\complex$, as follows,
\begin{align}\label{eq:multipath-steering}
a_{\phi,r,\rho}(h)\defin \add(h)+
\rho \ard(h)+
\rho \adr(h)
+
\rho^2 \arr(h),
\end{align}
where the components in the sum model the phase shift for each pair of Tx and Rx elements along the direct-direct~(DD) path to the scatterer located at $(\phi,\theta,r)$, the reflected-direct~(RD) path, and so on.
 Note that one can approximate
 $\add(h)=\ard(h)=a_{\text{Virt}}(\xi,r)$ where $\xi=[\phi, \theta]$ corresponds to the elevation $\theta$ of the direct path, even though there is an extra phase shift because of the difference in length of the DD and RD paths, and similarly  $\adr(h,r)=\arr(h,r)=a_{\text{Virt}}(\xi,r)$ where $\xi=[\phi, \theta']$  to the elevation $\theta'$ of the reflected (mirror) path, which is a function of the height and the distance of the scatterer. However, this approximation is not necessary for our method because exact numerical evaluations for any re-parametrization of the coordinates of the hypothetical target.

The coefficient $\rho$
 quantifies the phase shift and attenuation due to the interaction with the ground. Typical values of the reflection coefficient can be found for vertical and horizontal polarization in~\cite{EFK:93}. According to the reference, for low targets the phase shift $\arg(\rho)$ tends to be near $180^\circ$, and the attenuation factor $\absolute{\rho}$ is less than $1$. For the model~\eqref{eq:multipath-steering}, a good approximation of the unknown reflection coefficient $\rho$ is important to relate the height of the object to the super-position of the multi-path signals present in the measurement, and we describe a formulation to estimate it jointly with the height.
 
\paragraph*{Assumptions on range-Doppler bin} The range or range-Doppler bins of relevant detections are given. The considered ground-reflected multi-path contributions have a negligible influence on the estimation of range and Doppler for the case of small objects at far distances because the different contributions cannot be separated in the range or Doppler domains, and they appear as a single detection. This argument may not apply to high objects like bridges, especially when they are at close distances because the delays of each of the paths may appear in different range bins. The latter case can be addressed in multiple ways, e.g., extending the present sparse reconstruction models to include the range and Doppler dimensions~\cite{DC-YCE:18,CK-BS-SS-FR-RFHF-CW:18}, or complementing our algorithm with one of the methods described in the introduction for high targets. In principle, Doppler estimation is optional but is desirable because it provides an additional dimension for target separation and in some cases it brings an increase of SNR, e.g., when using a sequence of linearly modulated chirps. In what follows, if Doppler processing is performed previously to the angular estimation, then we assume when necessary that the Doppler frequency of the relevant Doppler bin is compensated for each pair of Tx and Rx elements and therefore assumed zero. 
  
  The estimation of azimuth and height can in principle be performed jointly but there is a curse of dimensionality in grid-based methods associated to the number of combinations of hypotheses, specially if high resolution is desired. Nonetheless, methods like BOMP can be decomposed and parallelized using multiple cores or GPUs.
 Therefore, we propose two strategies, a joint estimation of azimuth and height, and a sequential estimation procedure with two stages: First, azimuth is estimated for each detected rage-Doppler bin using a subset of Tx and Rx elements at the same height; Second, the height of the scatterers is estimated for each of the detected azimuth bins simultaneously using the measurements for all pairs of Tx and Rx elements. In each of the stages we combine array measurements across multiple points along the trajectory of the vehicle, cf.~Fig.~\ref{fig:diagram-car-trajectory}. We describe next this method.
  
 \subsection{Step 1: azimuth estimation}\label{subsec:azimuth-estimation}

 In the first stage, azimuth is estimated using a subset of coherent transmitters and receivers at the same height, cf.~Fig.~\ref{fig:geometries_2d}. Fig.~\ref{fig:geometries_2d} shows the Tx and Rx elements of exemplary sparse arrays with spatial diversity along the axes parallel and perpendicular to the ground. 
 %
%
 For a set of Tx elements and Rx elements, the set of antennas $\{p_i=[x_i, y_i, z]\tp\}_{i=1}^m$ has constant height, $z_i=z$, and the quantity $ n\tp p$ in~\eqref{eq:near-field-steering} can be written as
 \begin{align}\label{eq:far-field-phase}
 n\tp p =&\, \cos(\theta)\cos(\phi) x_i + \cos(\theta)\sin(\phi) y_i +  \sin(\theta) z\notag
 \\
= &\,u x_i  + v y_i  +  \sin(\theta) z ,
 \end{align}
 where $u\defin \cos(\theta)\cos(\phi)$ and $v\defin \cos(\theta)\sin(\phi)$ are the spatial frequencies for the variation of the phase along the depth and horizontal axes (so called because in the far-field these quantities determine the periodic variation of the phase measured at the antennas, but in the near-field model~\eqref{eq:near-field-steering} there is the additional quantity $\norm{p_i}^2/r^2$ in the square root). Note that the vertical frequency $w\defin  \sin(\theta)$ does not produce any phase variation because the sub-array is assumed at a constant height, i.e., the quantity $\psi\defin\sin(\theta) z$ contributes as an initial phase, for each source, common to all the antennas (approximately, since there is also the extra term $\norm{p_i}^2/r^2$ in the square root). 
 Note that if a quantity contributes as an initial phase, it can be factored out of the steering vector~\eqref{eq:near-field-steering} and neglected because the sparse reconstruction coefficients $x$ in~\eqref{eq:mininimization-group-sparse-reconstruction} absorb the phases common to all the antennas. However, if there are four paths in the contribution due to the elevation of the target, as explained in Section~\ref{subsec:multipath-steering-vector}, then this initial phase is not common to all the paths and cannot be factored out.
 
 This suggests three alternative strategies for azimuth estimation using progressively fewer assumptions:

 	\paragraph*{(i) Very low angles} If high targets are somehow filtered out, for very low targets one can use the approximation $\cos(\theta)\approx 1$ for $\theta\approx 0^\circ$, and therefore  $n\tp p =\cos(\phi) x_i  +  \sin(\phi)y_i  +  \psi$, so that it suffices a grid over azimuth angles in the FoV assuming $\theta=0$ and using as model the near-field steering vector~\eqref{eq:near-field-steering}.
  
	
	\paragraph*{(ii) Low angles with negligible ground reflections} In this case the contribution of the elevation to phase measurements renders necessary a grid of hypotheses for $u,v$ in the curve  $u^2+v^2=\cos^2(\theta)$ but since $\theta$ is unknown, one requires a grid in the circle $u^2+v^2\le\cos^2(\thetamax)$, or, expanding the domain, in the square $u,v\in[-\cos(\thetamax),\cos(\thetamax)]$. Nonetheless, because in this case there are no ground-reflections, it is still possible to use the near-field steering vector~\eqref{eq:near-field-steering} where the initial phase contribution $\psi\defin\sin(\theta) z$  is set to $0$ because is assumed factored out into the reconstruction coefficients (despite the term $\norm{p_i}^2/r^2$ in the square root).
	
		\paragraph*{(iii) Low angles with non-negligible ground reflections} In the case of multi-path, the initial phase cannot be factored out of the super-position of near-field steering vectors, and the sensing matrix is constructed evaluating the multipath model~\eqref{eq:multipath-steering} for combinations of hypotheses of azimuth in a fine grid $\{\phi_1,...,\phi_N\}\in[\phimin, \phimax]$ and hypotheses of height in a coarse segmentation $\{h_1,...,h_\nhprime\}\in[\hmin, \hmax]$, i.e., with $N>\nhprime$. 
		In this case it is unnecessary to select a subset of antennas at the same height, and indeed all antennas can be used. The trade-off between desired height resolution and the computation and memory cost due to the dense grid in each dimension (number of columns of the sensing matrix) and number of antennas (rows), depends on whether it is desired to have a second stage where $1$-dimensional height estimation is performed for the detected azimuth bins.

 Note that the curvature on the $x$-$y$ plane of a conformal array would add further variability of the phase, which can be beneficial for azimuth estimation. In the cases~(i) and~(iii) this is leveraged without additional computation complexity but in case~(ii) it could also be admissible because the grid over~$u$ can be coarser than for~$v$ if the aperture on depth $\max_i \{x_i\}-\min_i \{x_i\}$ is small compared to the horizontal aperture $\max_i \{y_i\}-\min_i \{y_i\}$ (i.e., almost planar), because then the expected resolution is not as high for the estimation of $u$ as it is for $v$. Indeed, if depth is constant,  $x_i=x$ (no curvature), then in case~(ii) it is only required a grid over $v\in[-\cos(\thetamax),\cos(\thetamax)]$, and the term $\psi'\defin \cos(\theta)\cos(\phi) x +  \sin(\theta) z$  contributes as an initial phase common to all the antennas for each source. Since case (ii) assumes no multi-path, then this initial phase  can be approximately factored out of the near-field steering vector~\eqref{eq:near-field-steering} into the reconstruction coefficients.

With these assumptions, the sparse reconstruction problem for azimuth estimation is given by
 $\ysubhor{r}\approx \Asubhor{r} \xsub{r}$
 where $\ysubhor{r}$ contains the complex spectral values for the relevant range-Doppler bin associated to a  subset of transmitters and receivers, each at the same height in cases~(i) and~(ii), or possibly all the antennas in case~(iii).
 
 The sensing matrix for azimuth estimation  $\Asubhor{r}$ is defined according to the cases (i)-(iii) described above as follows:
 \begin{enumerate}[(i)]
 	\item As in~\eqref{eq:sensing-matrix-azimuth} for the near-field steering vector~\eqref{eq:near-field-steering}-\eqref{eq:virtual-steering} evaluated at $\{\phi_1,...,\phi_N\}\subset[\phimin, \phimax]$ and setting the $\theta=0$.
 	
 	\item Stacking as columns the  near-field steering vector~\eqref{eq:near-field-steering}-\eqref{eq:virtual-steering} re-parametrized for combinations of hypotheses $\{(u_1,v_1),...,(u_N,v_N)\}\in[-\cos(\thetamax),\cos(\thetamax)]$ and setting the term $\psi=\sin(\theta) z$ in~\eqref{eq:far-field-phase} to $0$.
 		
 \item  Stacking as columns the multi-path steering vector~\eqref{eq:multipath-steering} evaluated in combinations of hypotheses of azimuth in a fine grid $\{\phi_1,...,\phi_N\}\in[\phimin, \phimax]$ and hypotheses of height in a coarse segmentation $\{h_1,...,h_\nhprime\}\in[\hmin, \hmax]$. The treatment for the reflection coefficient is described in Section~\ref{subsec:height-estimation}.

 \end{enumerate}

The sparsity-inducing penalty for the reconstruction coefficients $\xsub{r}$ is defined as in~\eqref{eq:group-sparsity-constraint}, as a function of one or more 
mutually incoherent array measurements. The incoherent combination of measurements along the trajectory or from multiple sensors is briefly described in Section~\ref{subsec:range-sensor-fusion}.

 The azimuth map $\xazimap$ is defined in a similar fashion to~\eqref{eq:doa-reconstruction-map} taking the average or the maximum over all reconstruction coefficients corresponding to each azimuth hypothesis. By thresholding the  entries of $\xazimap$, we obtain in case (i) or (iii) the associated azimuth bins of detected targets $\{\phi_1,...,\phi_\naz\}$, and in case~(ii) the frequency bins $\{(u_1,v_1),...,(u_\naz,v_\naz)\}$. 
 If the density of the grid of height hypotheses in case~(iii) is high, and all antenna measurements are used, this case can be regarded as a joint estimation of azimuth and height.

\subsection{Step 2: Height estimation}\label{subsec:height-estimation}

Once the azimuth bins $\{\phi_1,...,\phi_\naz\}$ or spatial frequencies $\{(u_1,v_1),...,(u_\naz,v_\naz)\}$  of candidate targets are estimated, the second stage is the estimation of the height of scatterers at each azimuth bin. The vector of data points $\ysub{r}$ used for this estimation task corresponds to the complex spectral values for the relevant range-Doppler bin associated to the full set of transmitters and receivers available, cf.~Fig.~\ref{fig:geometries_2d}. That is, whereas for azimuth estimation we assumed a measurement acquisition from  Tx and Rx elements at the same height, this tasks demands spatial diversity both in the vertical axis, $z$~axis (to distinguish hypotheses of height) and in the horizontal and/or depth axes, $x$ or $y$~axes (to discriminate between the hypotheses of height for targets located at different azimuth angles).

The sparse reconstruction problem is $\ysub{r} \approx \Asub{r} \xsub{r}$ where
$\Asub{r}$ contains the multi-path steering vector~\eqref{eq:multipath-steering} for each hypothesis of height $\{h_1,...,h_\nh\}$
for each of the detected azimuth bins $\{\phi_1,...,\phi_\naz\}$ and for each hypothesis of the reflection coefficient $\{\rho_1,...,\rho_\nrho\}$, namely,
\begin{align*}
\Asub{r}\defin[A_{\phi_1,r,\rho_1}\cdots A_{\phi_\naz,r,\rho_1} \cdots A_{\phi_\naz, r,\rho_{\nrho}}]
\in\complex^{M\times \nh\naz\nrho}
\end{align*}
with
\begin{align*}
A_{\phi,r,\rho}\defin[a_{\phi,r,\rho}(h_1) \cdots a_{\phi,r,\rho}(h_\nh) ]\in\complex^{M\times \nh} .
\end{align*}
The group-sparse penalty on the reconstruction coefficients~$\xsub{r}$ has label vector (cf.~\eqref{eq:label_vector}) according to one of the following two options,
\begin{subequations}\label{eq:label-vector-height}
\begin{align}
v_a=&\,\ones_{\nrho}\otimes[1,...,\nh\naz] ,
\\
v_b=&\,[1,...,\nh\naz\nrho] ,
\end{align}
\end{subequations}
where $\ones_{\nrho}$ is the vector of ones of length $\nrho$.
In option~(a) $v_a$ codifies a penalty on the number of height bins associated to each detected azimuth bin, but we allow a superposition of multi-path steering vectors~\eqref{eq:multipath-steering} for several values of the reflection coefficient, without penalizing sparsity in this selection, whereas in option~(b) we encourage using the partition $v_b$ the reflection coefficient to take the smallest possible number of values. The desired option can be judged empirically using real measurements.

The height map $\xheightmap_{\phi_j}(i_h)\in\real^N$ for each azimuth bin $\phi_j$, is computed for each height hypothesis~$h$ by averaging or taking the maximum over all hypotheses of~$\rho$,
\begin{align}\label{eq:height-map}
\xheightmap_{\phi_j}(i_h)\defin\frac{1}{ \nrho} \ones_{\nrho}\tp  \xsub{r}[ i_{h, \phi_j} : N \naz:  N\naz \nrho], 
\end{align}
for $i_{h, \phi_j} \in\until{N\naz}$, where $\ones_{ \nrho}$ is the vector of ones of size $\nrho\times 1$. (This is a pooling operation or projection onto height hypotheses for each azimuth bin, by merging information about the nuance parameter $\rho$, although if required, the parameter $\rho$ can also be estimated from the above reconstruction.)

Some observations are the following:
(a) In the sensing matrix $\Asub{r}$, we have included as columns the model~\eqref{eq:multipath-steering} with several values of $\rho$, in particular using 
a coarse grid of attenuation values in the interval $[0,1]$ with an off-grid error of $0.1$. Thus, the number of columns, $\nh\naz\nrho$, is not prohibitively large because $\naz$ and $\nrho$ are small.
(b) The above sensing matrix illustrates the possibility of including multiple complementary or competitive models for signal reconstruction, thus avoiding the computational cost of model testing as in~\cite{FE-MW-PH:17}. 
The columns can be interpreted as features based on physical models with some uncertain parameters like the reflection coefficient, and therefore these models are a simple alternative to blind calibration and more general dictionary learning which require a larger number of snapshots.



\subsection{Incoherent fusion across filtering interval}\label{subsec:range-sensor-fusion}

The measurements at different points of the trajectory can be processed jointly with the group-sparse model of Section~\ref{subsec:gs-reconstruction}. We describe it formally and then state the required assumption of hypothesis consistency. The set of sparse reconstructions for 
azimuth and height estimation in Sections~\ref{subsec:azimuth-estimation} and \ref{subsec:height-estimation} can be written as
\begin{align}\label{eq:sparse-system-all-ranges}
\{\ysub{r_i} \approx \Asub{r_i} \xsub{r_i}\}_{i=1}^{I_j} ,
\end{align}
where $r$ is the range (or range-Doppler) associated to a given detection from different points along the trajectory of the vehicle $\{p_j^{(1)},...,p_j^{(I_j)}\}$ for filtering interval $j$, cf.~Fig.~\ref{fig:diagram-car-trajectory}. 

The sparsity constraint or penalty is responsible for the coupling between the sparse reconstructions at the different points in the filtering interval. To use the notation in Section~\ref{subsec:gs-general-partition} that is more general than in~Section~\ref{subsec:gs-reconstruction},
we re-write $y\defin[\ysub{r_1}\tp \cdots \ysub{r_{I_j}}\tp]\tp$, $A\defin\diag(\Asub{r_1}, \cdots, \Asub{r_{I_j}})$ (i.e., in block-diagonal form) and $x\defin[\xsub{r_1}\tp \cdots \xsub{r_{I_j}}\tp]\tp$,
and extend the group partition to the case of multiple points along the trajectory by defining the label vectors,
\begin{subequations}\label{eq:label-vector-range-fusion-height}
\begin{align}
v_a=&\,\ones_{I_j\nrho}\ones_{\nrho}\otimes[1,...,\nh\naz] ,
\\
v_b=&\,\ones_{I_j}\otimes[1,...,\nh\naz\nrho] ,
\end{align}
\end{subequations}
where $I_j$ is the number of incoherent measurements along the trajectory that are coupled by the group sparsity constraint, i.e., that are encouraged to be fit by the same height and azimuth hypotheses (up to model corrections based on odometry systems). In option~(a), in contrast with option~(b), the reflection coefficient can be distinct at each point because we do not impose the selection of these values to be sparse.
An identical formal model is used to combine measurements from several mutually incoherent sensors in the vehicle. We refer to both options as \textit{group-sparse}~(``GS") \textit{range fusion}.

The height map for this case is computed from~\eqref{eq:height-map} by further averaging the reconstructions at different points of the trajectory, and optionally for two or more mutually incoherent sensor measurements. To have a standardized way of thresholding the height map, we suggest to normalize the signals at each distance to the obstacle $\{\ysub{r_i}\}_{i=1}^I$ and similarly the columns of the sensing matrices $\{\Asub{r_i}\}_{i=1}^I$. (We have studied the benefit of not normalizing column-wise across hypotheses of height to keep the information of the power of the destructive and constructive interference but results were not improved.)
Then, according to the SNR at each distance, one can also include a weighting factor in the average, e.g., proportional to inverse distance to detection.

To study the benefit of coupling the reconstructions at different points of the trajectory via the group-sparsity constraint for the partition~\eqref{eq:label-vector-range-fusion-height}, we compare in the next section the performance with the alternative approach of solving the sparse reconstructions~\eqref{eq:sparse-system-all-ranges} individually at each distance to the obstacle with the group partition~\eqref{eq:label-vector-height}. We call this approach \textit{step-by-step}~(``SbyS") \textit{range fusion}, versus the~``GS" approach. The height map in~\eqref{eq:height-map} is averaged over the points in the trajectory in both cases, but the group-sparsity constraint~\eqref{eq:label-vector-range-fusion-height} that encourages the hypothesis of height to be the same across distances is only included in the ``GS" approach.


\subsection{Data and hypothesis consistency} 
The formulations in the previous section for angular and height estimation along the filtering interval of the trajectory of the vehicle need the following assumptions.

\paragraph*{Data consistency}The assumption of data consistency says that the data vector $\ysub{r_i}$ (for the antenna measurements) corresponds to the range-Doppler bin of the same target of interest for all points along the filtering interval, $i\in\until{I}$. That is, we assume that the range-Doppler bin may have changed along the filtering interval, and those detections are tracked or associated with one another. Inertial systems and Doppler measurements can be used for this task. 

\paragraph*{Hypothesis consistency} For azimuth or height estimation, this condition states that the entries of the sparse reconstruction~$\xsub{r_i}$ and the associated columns of the sensing matrices~$\Asub{r_i}$  correspond to the same hypotheses formulated for the same detection(s). 

Hypothesis consistency  for azimuth estimation:
\begin{enumerate}[(1)]
	\item	The range-Doppler bin for which the hypotheses of azimuth are formulated at different points along the trajectory, correspond to the same detected target, or detection (in concordance with the aforementioned data consistency).
	
	\item	The columns of the sensing matrix correspond to the same hypotheses of azimuth.
\end{enumerate}

Hypothesis consistency for height estimation:

\begin{enumerate}[(3)]
	\item	The previously estimated azimuth bins, possibly varying along the trajectory (for a single range-Doppler bin) for which hypotheses of height are formulated, correspond to the same detected targets.
	
	\item	The columns of the sensing matrix correspond to the same hypotheses of height.
\end{enumerate}

\paragraph*{Discussion of applicability}
 Condition~(1) for azimuth is trivial if data consistency is achieved.
For height estimation, condition~(3) can be satisfied assuming that the same azimuth bins are present in the signals~$\{\ysub{r_i}\}_{i=1}^I$ along a sufficiently short filtering interval.
 Note that the sensing matrix for height estimation for each point along the trajectory can be constructed in terms of the previously estimated azimuth bins, even if they change from point to point, provided that they correspond to the same target.

\begin{figure}[]
	\includegraphics[width=0.99\linewidth]{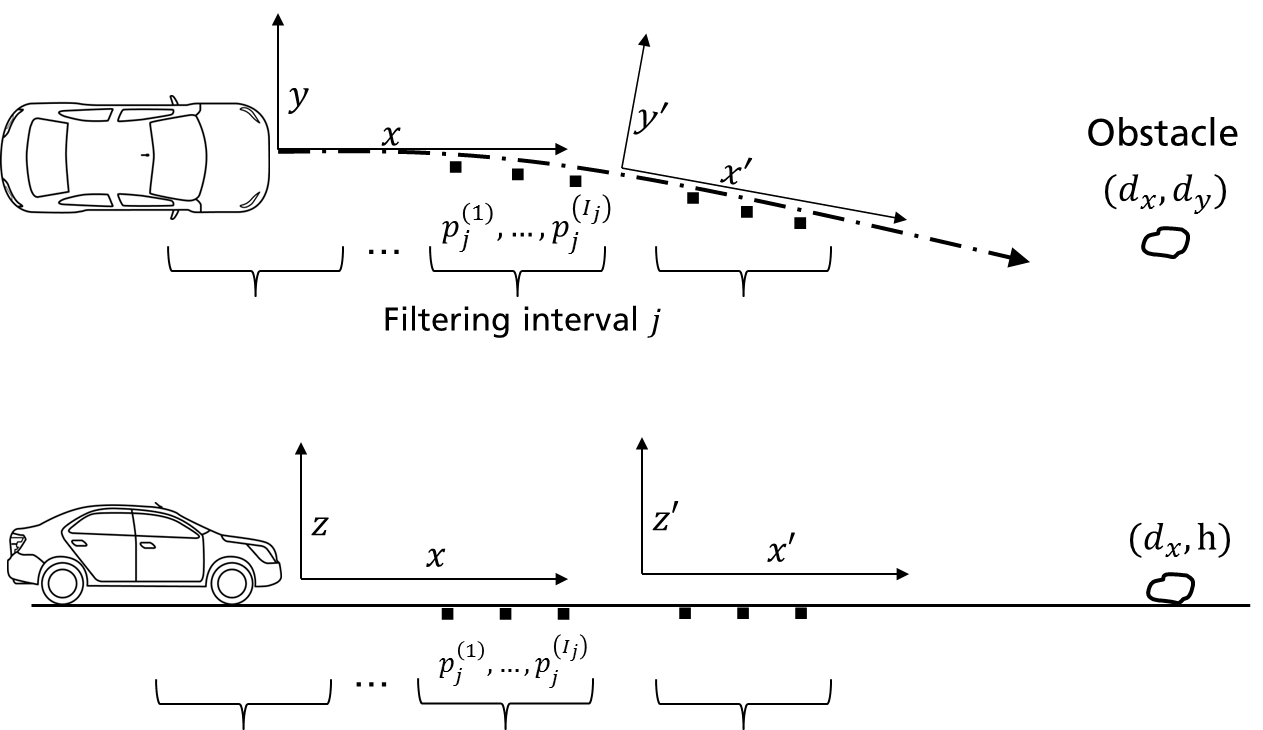}
	
	\caption{Measurement acquisition along trajectory of the vehicle. Measurements are collected constantly, and the Range-Doppler map and CFAR are computed at every point. The proposed algorithm for azimuth and height estimation is applied to candidate detections jointly over a filtering interval of points $\{p_j^{(1)},...,p_j^{(I_j)}\}$.	
	}
	\label{fig:diagram-car-trajectory}
\end{figure}

Condition~(2) and (4) can be satisfied, for azimuth or height estimation, if the coordinates of azimuth and/or height of the target in the frame of reference that moves with the vehicle have not changed, cf.~Fig.~\ref{fig:diagram-car-trajectory}. In~Fig.~\ref{fig:diagram-car-trajectory}~(top) we observe that the azimuth or transversal Cartesian coordinate can change along the trajectory of the vehicle. An alternative in this case is including a correction in the hypotheses of the sensing matrix, e.g., if the motion of the vehicle is known thanks to odometry systems. In~Fig.~\ref{fig:diagram-car-trajectory}~(bottom) we observe that for a flat road, the height of the object (as opposed to the elevation angle) remains constant regardless of the trajectory. 

%

%

\section{Simulation results}\label{sec-simulation results}

In this section we analyze the proposed algorithms using Monte Carlo simulations for $1$ and $2$~point scatterers. First we study the case where scatterers are in the same azimuth bin, for arrays that have only vertical diversity in antenna positions. In the next section we consider a 2-dimensional array for sequential azimuth and height estimation where the targets are not necessarily in the same azimuth bin, but rather it is required to resolve them in both dimensions.

In particular, we consider three aspects: i) comparison of sparse reconstruction using one radar array or two mutually incoherent radar arrays located in the bumper and in the top, versus  methods that use the power envelope of the interference signal, ii) design recommendations for aperture, height of the sensor,  geometry, and number of antennas, and iii) the performance of the sequential method for azimuth and height estimation using several arrangements of antennas with vertical and horizontal diversity.

\subsection{Comparison of sparse reconstruction with MUSIC and Burg method}

%
\begin{figure*}[]
	\begin{subfigure}[b]{0.48\textwidth}
		\includegraphics[width=\textwidth]{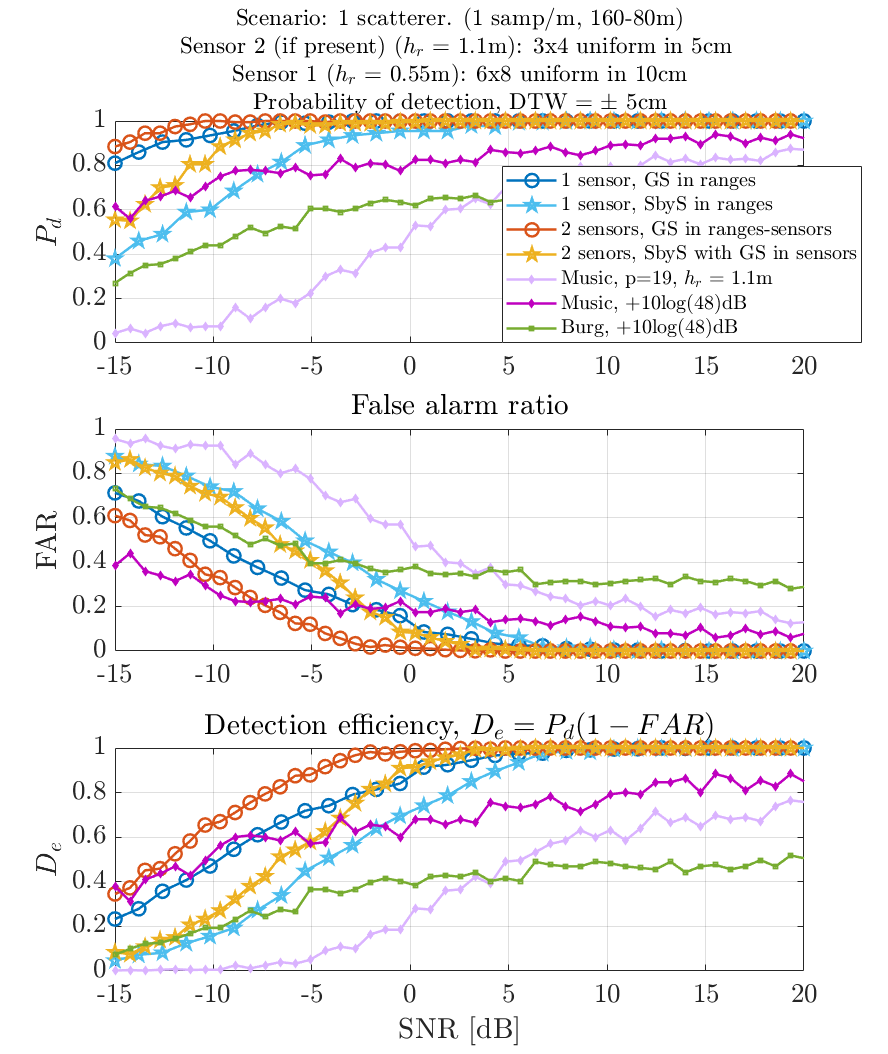}
		\subcaption{Scenario with 1 scatterer.}
		\label{fig:cs_6x8_1tar}
	\end{subfigure}
	\begin{subfigure}[b]{0.48\textwidth}
		\includegraphics[width=\textwidth]{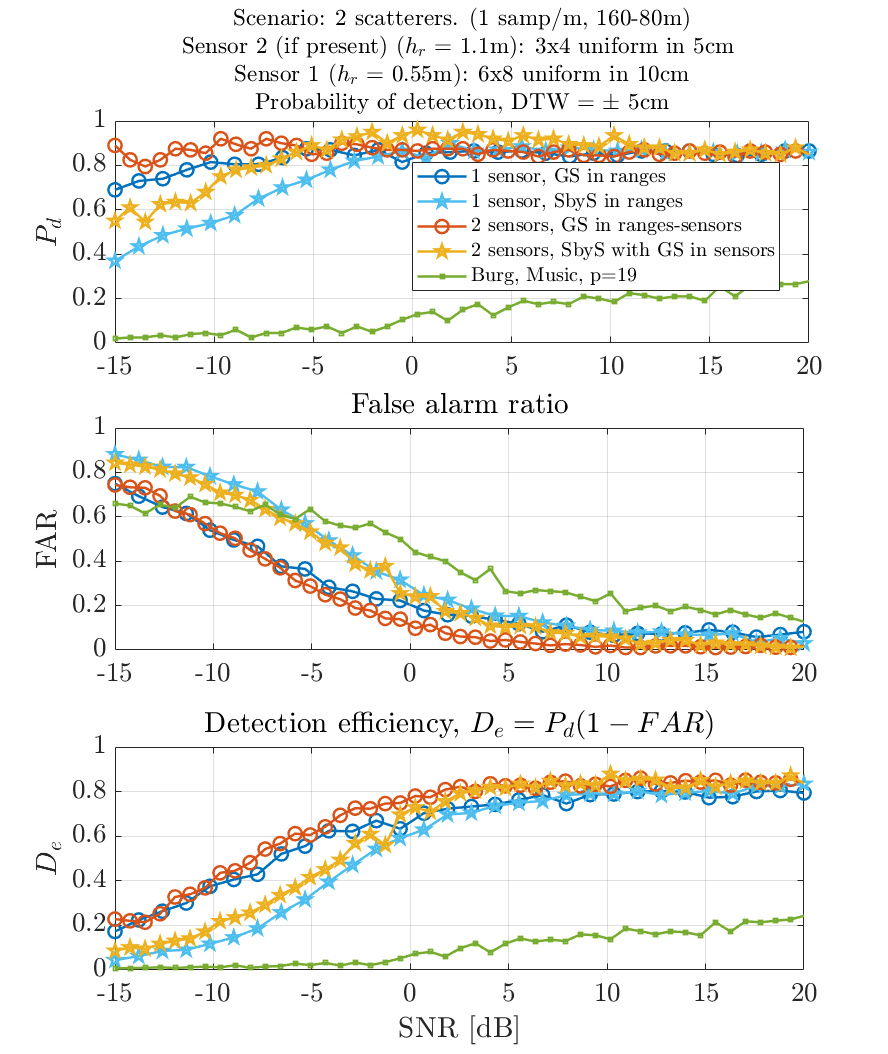}
		\subcaption{Scenario with 2 scatterers.
		}
		\label{fig:cs_6x8_2tar}
	\end{subfigure}
	\caption{Comparison using synthetic measurements. The  SNR is quantified as signal power to noise-variance ratio per antenna. 
		The comparison includes the case of 1 sensor of 6x8 elements uniformly spaced in 10 cm, located in the bumper at a height of 0.55m, and the case of adding a second incoherent sensor of 3x4 elements uniformly spaced in 5cm at the top of the vehicle at a height of 1.1 m. 
		The group-sparse (``GS") reconstruction over distances is done using a filtering interval of 8m, and there are in total 10 such intervals over which an average of the height maps is performed. Step by step (``SbyS") performs an average of the individual reconstructions before thresholding. Our method estimates the reflection attenuation coefficient simultaneously with the height. We have added for comparison MUSIC and Burg using an antenna placed at $1.1$~m that measures the interference pattern over the same interval of $160-80$ m to the obstacle.
	}
	\label{fig:cs_6x8}
\end{figure*}

The performance of our algorithm is shown in Fig.~\ref{fig:cs_6x8} for $1$ and $2$ mutually incoherent arrays. Sensor~1 has 6x8 elements uniformly spaced in $10$ cm and is located in the bumper at a height of $0.55$ m, and sensor~2 (incoherent with the first) has 3x4 elements uniformly spaced in $5$ cm and is on the top of the vehicle at a height of $1.1$ m. 
 

We show the scenarios of~$1$ and~$2$ point scatterers placed randomly at a height between $0.1$ m and $1.35$ m, with a random separation, and at the same azimuth bin. The statistical performance metrics are computed over $200$ Monte Carlo realizations. The metrics we use are  Probability of detection ($P_d$), False alarm ratio (FAR) and detection efficiency ($D_e$) (cf.~\cite{MGH-DMN-CG-RS:18}) for a detection window (DTW) of $\pm5$ cm around the highest scatterer only. (That is, we only consider false alarms if they are outside the DTW above the highest scatterer.) The DTW is defined in~Section~\ref{subsec:gs-reconstruction}.

We have exemplified two versions of our algorithms:

\begin{enumerate}[(i)]
	\item  Step by step (``SbyS") estimation of height, where at each point in the trajectory sparse reconstruction is performed with phase and amplitude measurements of a single array (or optionally $2$ mutually incoherent arrays);
	\item Group-sparse (``GS") reconstruction of height using phase measurements across multiple points in a filtering interval (cf.~Fig.~\ref{fig:diagram-car-trajectory}).
\end{enumerate}

The group-sparse reconstruction (``GS") over distances is done using a filtering interval of $8$ m, and there are in total $10$ such intervals, from $160$ m to $80$ m to the target.
 In the shown example with one sensor, the reflection attenuation coefficient is estimated jointly with the height.



For comparison we have added an algorithm that uses the principle of periodicity of the envelope of the ground-reflected  interference signal in the scale of inverse range to the target~\cite[pp. 451]{SMK:81}. 
This principle has been used for bridge identification~\cite{FD-JK-FS-JD-KD:11}, where the height of a single scatterer is related to the peaks of the power spectral density, and the latter is estimated with the FFT but we use instead the parametric auto-regression Burg method or MUSIC, which are methods that require fewer samples than the FFT. It is important to remark that for two or more scatterers spectral methods based on the interference pattern fail because in this model there is nonlinear superposition of target echoes.
%

In the comparison of Fig.~\ref{fig:cs_6x8}, we have assumed that Burg and MUSIC have been applied using a single antenna placed at $1.1$ m. Note that this method works better for a higher antenna because the number of cycles of the interference pattern grows with the height of the target and the height of the radar.

We have also been generous with respect to MUSIC and Burg adjusting the SNR for the application of these methods in relation to the ideal SNR gain corresponding to the number of virtual pairs of Tx and Rx used for sparse reconstruction, in this case adding $10 \log(48)$ dB because our algorithm uses a coherent block of 6x8 antennas.
We note that MUSIC performs better than Burg (for both SNR scales). There is a parameter that has been approximately tuned for MUSIC and Burg, namely, the subspace dimension of the signal in MUSIC and the order of auto-regression in Burg. 
In the ideal case of an SNR gain corresponding to an equivalent number of 6x8 antennas, we observe that MUSIC performs as well as a 6x8 sensor only at very low SNR, while the proposed algorithm is still superior for the rest of SNR values even with this generous scale that benefits MUSIC. We observe that our methods achieve remarkable detection efficiency and are valid for one and more scatterers. In this respect, we remark that we have used BOMP with sparsity number $2$, because we do not expect more than 2 scatterers in the same range-Doppler and azimuth bin, but it also works setting a higher threshold as stopping criterion.

In Table~\ref{table:design-recommendations} we summarize some results (not shown) of a parametric analysis studying several design choices. 
\begin{table}[t]
	\centering
	\caption{Summary of design recommendations}
	\label{table:design-recommendations}
	\begin{tabular}{ p{1.6cm}| p{2.9cm} | p{2.9cm} }
		\hline
		\noalign{\vskip 1pt} 
		Aspect & Sparse reconstruction & PSD methods
		\\[1.1pt]
		\hline
		\hline
		\noalign{\vskip 1pt} 
		Aperture&
		Most relevant aspect for thin FoV & 
	NA
		\\[2.3pt]
		Height of sensor&
	Not so relevant & 
	The \textbf{highest the better}: $\#$~of cycles of interference envelope increases
		\\[2.3pt]
		Geometry&
		Important only for wide FoV & 
		NA
		\\[2.3pt]
		$\#$~Antennas&
		Relevant at low SNR & 
		Increases SNR
			\\[2.3pt]
	$\#$~Incoherent arrays (height)&
		Different heights add diversity in SNR of constructive/destructive interference signal
		 & 
		As for $\#$ Antennas, because is always incoherent
			\\[2.3pt]
		$\#$~Incoherent arrays (azimuth)&
	 Optimization of antenna positions for sidelobe averaging between incoherent apertures~\cite{DMN-MGH-RS-ACS:19}
		& 
		NA
		\\
		\hline
		\hline
		\noalign{\vskip 2.2pt}
	\end{tabular}
\end{table}

\subsection{Sequential processing for azimuth and height estimation}

\begin{figure*}[]
	\begin{subfigure}[b]{0.50\textwidth}
		\includegraphics[width=\textwidth]{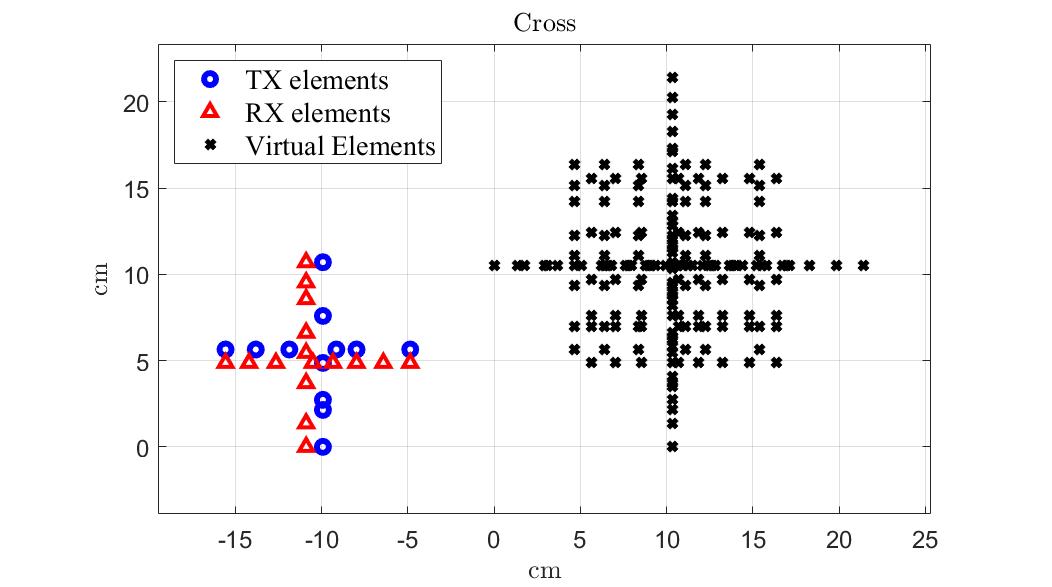}
		\label{fig:cross}
	\end{subfigure}
	\begin{subfigure}[b]{0.50\textwidth}
		\includegraphics[width=\textwidth]{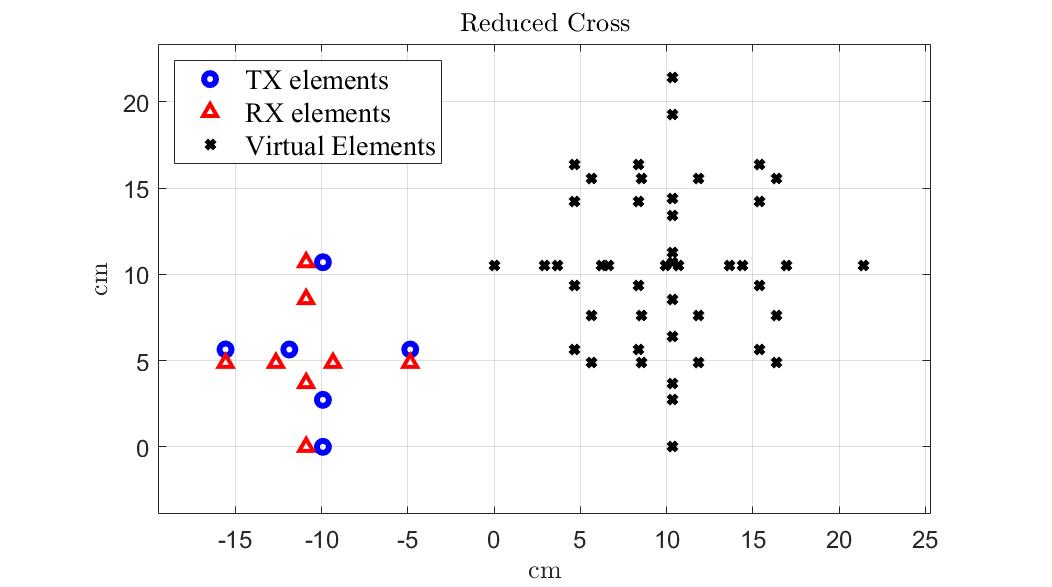}
		\label{fig:cross_reduced}
	\end{subfigure}
	
	\begin{subfigure}[b]{0.50\textwidth}
		\includegraphics[width=\textwidth]{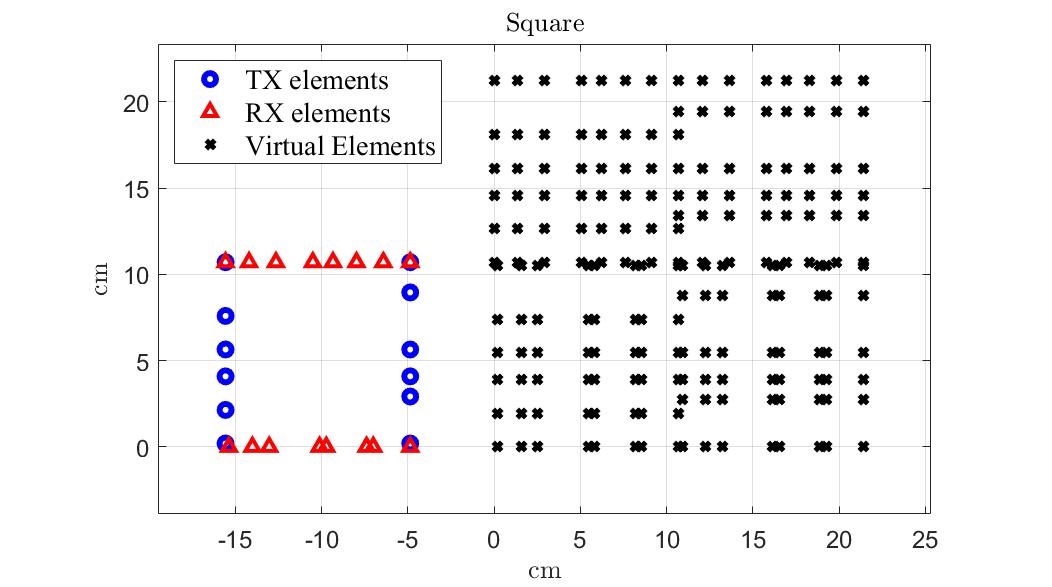}
		\label{fig:square}
	\end{subfigure}
	\begin{subfigure}[b]{0.50\textwidth}
		\includegraphics[width=\textwidth]{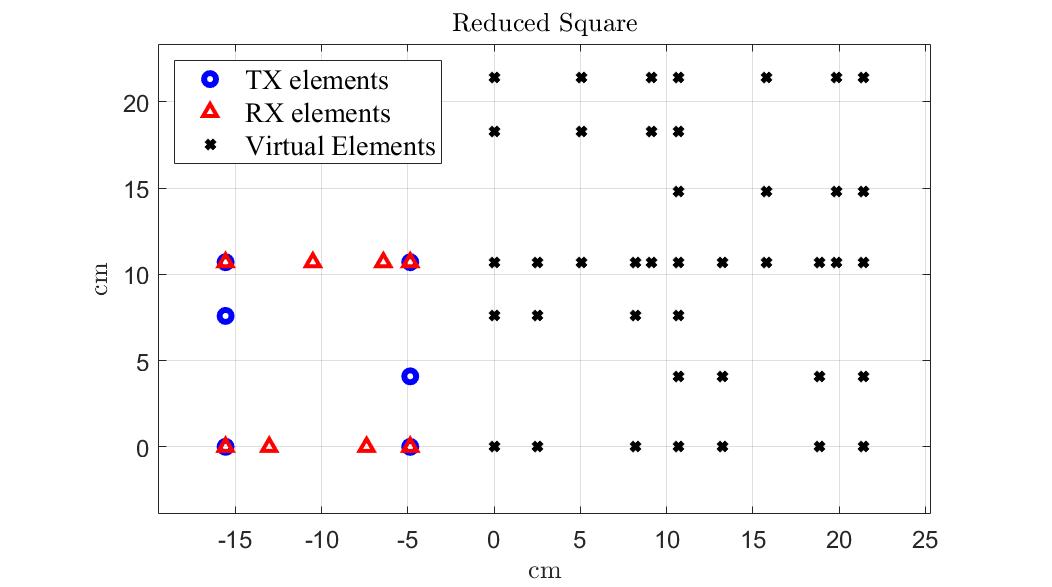}
		\label{fig:square_reduced}
	\end{subfigure}
	\caption{Geometries of 2-dimensional arrays considered for sequential azimuth and height estimation. The antenna positions are heuristically chosen to increase the surface of the virtual array. These arrays can be improved optimizing rigorously the sidelobe level and the mainlobe width in the joint parameters of azimuth and elevation.
	}
	\label{fig:geometries_2d}
\end{figure*}

Here we study the effect of the geometry and number of elements of 2-dimensional arrays on the performance of sequential azimuth and elevation estimation described in Section~\ref{sec-height-estimation-model} combined with step-by-step range processing.  For this analysis we propose two types of 2-dimensional geometries with different number of elements placed within the typical maximum space available of $10$ to $11$cm in the horizontal and vertical axes (cf.~Fig.~\ref{fig:geometries_2d}). The particular selection of the positions for the Tx and Rx channels has been motivated by the 2-dimensional distribution of virtual elements. The following educated guesses improve the diversity in the number and position of virtual elements for each array configuration. However, there is still room for improvement in these designs because in these examples we have not employed any dedicated technique for array design. In addition, here we only represent the step-by-step (``SbyS") approach for fusion of measurements along the trajectory, which in Fig.~\ref{fig:cs_6x8} is shown to be inferior to the approach of group-sparse (``GS") fusion.

\begin{figure*}[]
	\hspace*{1.5cm}
	\begin{subfigure}[b]{0.40\linewidth}
		\includegraphics[width=\textwidth]{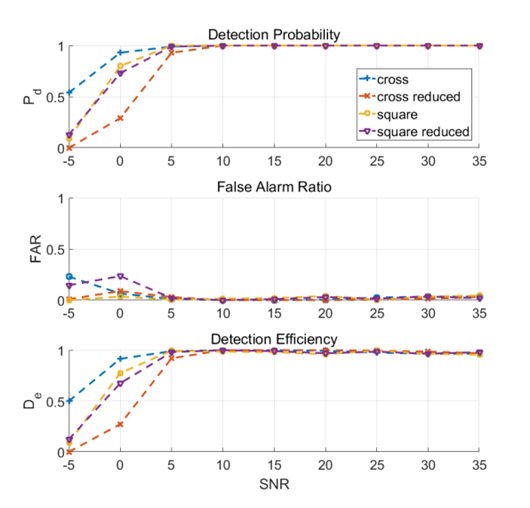}
		\subcaption{Scenario with 1 scatterer.}
		\label{fig:cs_2d_1tar}
	\end{subfigure}
	\begin{subfigure}[b]{0.40\linewidth}
		\includegraphics[width=\textwidth]{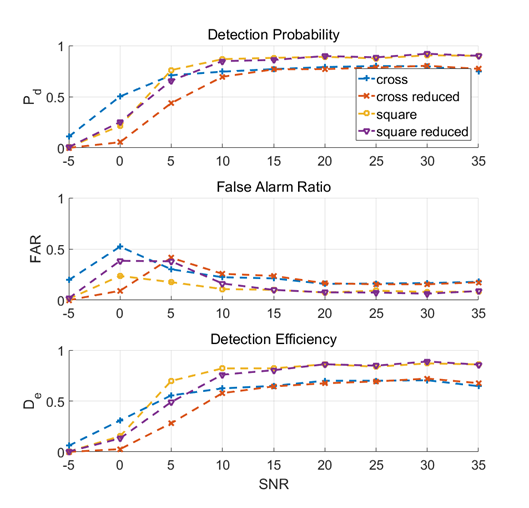}
		\subcaption{Scenario with 2 scatterers.
		}
		\label{fig:cs_2d_2tar}
	\end{subfigure}
	\caption{Performance of sequential estimation of azimuth and height using the 2-dimensional arrays in~Fig.~\ref{fig:geometries_2d}. We use the same metrics as in Fig.~\ref{fig:cs_6x8} but computed with respect to the two targets (as opposed to computing them with respect to the highest target) with a detection window of $\pm 4cm$. The targets are randomly generated in each of 200 Monte Carlo realizations with arbitrary angular separation in azimuth and elevation. These are preliminary results where we have applied only the step-by-step (``SbyS") approach for fusion of measurements along the trajectory, and the probability of detection and false alarms can be further improved with optimization of antenna positions.
	}
	\label{fig:cs_2d}
\end{figure*}


The achieved results for one and two targets with $200$ Monte Carlos for each SNR value, and a detection window of $\pm 4$cm are presented in Fig.~\ref{fig:cs_2d}. 
With regards to the kind of geometry, we observe that for one target, the performance of the cross design in Fig.~\ref{fig:geometries_2d} outperforms the other designs at low SNR. One possible reason is that for a single detection, the spatial diversity in 2 dimensions is not so important, while this array has a higher number of virtual elements in each 1-dimensional axis, which is beneficial for the sequential method for azimuth and height estimation at low SNR. 
On the other hand, when there are two detections, the square designs show the best performance because they have a more diversity in the 2-dimensional virtual aperture; this is beneficial in the second stage of height estimation because this array can then discriminate and resolve better the heights of two targets. 

When we compare the 12x16 arrays with their 6x8 counterparts, we see that the number of elements affects the performance at low-medium SNR, while the aperture determines the best achievable performance. Interestingly, this maximum performance is identical for each type of geometry independently of the number of elements.

\section{Conclusions and future work}\label{sec-conclusions}

We have proposed a method for sequential estimation of azimuth and height of low objects using group-sparse reconstruction using a multi-path model with unknown reflection coefficient. The method allows incoherent combination of measurements along the trajectory and optionally from two or more mutually incoherent array sensors. We have shown the performance with simulated signals, comparing it with conventional power spectral methods that use the periodicity of the interference pattern. Current work includes testing the algorithm with radar measurements and comparing the performance with other state-of-the-art algorithms.
 %
We have shown several advantages of the sparse reconstruction framework, like the possibility of including in the sensing matrix multiple competing models, e.g., values for the reflection coefficient, to deal with uncertainty, and we have also provided design recommendations based on parametric simulations.
%
%
%

There are many open questions, notably
dealing with a non-flat world, e.g., filtering out road-reflected signals, or with objects that produce self reflections or weak reflections. 
In this regard, we have been challenged to use I-Q raw data, or signals in frequency domain before CFAR, but it would be intractable to apply our method to every range-Doppler bin so the problem remains of how to combine adaptively the detection and the sparse reconstruction, and combining the group-sparse reconstruction with priors and tracking, e.g., using the Earth-mover distance.
Another open question is the application of group-sparse regularization in more general domains, e.g., to combine radar data with different polarizations, and to use group-sparsity for classification using neural nets.

We have observed that our models yield improved accuracy and reduced false alarms when applying group-sparse regularization to the reconstruction coefficients of signals measured at multiple points along the trajectory, which is an improvement over the state-of-the-art for low SNR at least for simulated data of specular reflections with unknown reflection coefficient. Even if further investigation uncovers more realistic multipath models, we believe that the techniques shown here remain valid. We also observe the benefit of using multiple mutually incoherent arrays over one array, effect that can be further amplified, particularly for a wide field of view, by optimizing the antenna positions using metrics that enhance group sparse reconstruction, technique that we have demonstrated in parallel work.
%

%

\section*{Acknowledgment}

This work has been supported by Audi AG in the period Aug. 2017 - Feb. 2018, and we warmly thank Michael Schwenkert and Niels Koch from Audi AG, together with other anonymous persons, for championing this effort and
for their technical feedback. We also thank Christian Greiff and Fabio Giovanneschi for their camaraderie. Our gratitude also goes to our department head, Stefan Br\"uggenwirth.




\end{document}